# Negative Differential Resistance, Memory and Reconfigurable Logic Functions based on Monolayer Devices derived from Gold Nanoparticles Functionalized with Electro-polymerizable Thiophene-EDOT Units

T. Zhang[1], D. Guérin[1], F. Alibart[1], D. Vuillaume[1], K. Lmimouni[1], S. Lenfant[1*]

A. Yassin[2], M. Oçafrain[2], P. Blanchard[2], J. Roncali[2]

[1] Institute of Electronics Microelectronics and Nanotechnology (IEMN), CNRS, University of Lille, Avenue Poincaré, 59652 Villeneuve d'Ascq, France

[2] MOLTECH-Anjou, CNRS, University of Angers, 2 Bd. Lavoisier, Angers, 49045, France

**ABSTRACT**: We report on hybrid memristive devices made of a network of gold nanoparticles (10 nm diameter) functionalized by tailored 3,4-(ethylenedioxy)thiophene (TEDOT) molecules, deposited between two planar electrodes with nanometer and micrometer gaps (100 nm to 10 μm apart), and electro-polymerized in situ to form a monolayer film of conjugated polymer with embedded gold nanoparticles (AuNPs). Electrical properties of these films exhibit two interesting behaviors: (i) a NDR (negative differential resistance) behavior with a peak/valley ratio up to 17; and (ii) a memory behavior with an ON/OFF current ratio of about $10^3$ to $10^4$. A careful study of the switching dynamics and programming voltage window is conducted demonstrating a non-volatile memory. The data retention of the "ON" and "OFF" states is stable (tested up to 24h), well controlled by the voltage and preserved when repeating the switching cycles (800 in this study). We demonstrate reconfigurable Boolean functions in multi-terminal connected NP/molecule devices.

**INTRODUCTION:** Resistive organic memories have emerged as promising candidates for hardware implementation of circuitries in organic electronics. Several groups have demonstrated the resistive switching of hybrid organic- metallic nanoparticles materials, and many materials (organic, inorganic, and composite) have been used to demonstrate resistive switching.[1] Organic devices have the advantages of low fabrication cost, simple process, low weight, tunable material and high mechanical flexibility, and are quite promising for the research of the next generation RRAMs devices with a high density, multilevel storage and high flexibility. In organic electronics, the first was introduced by Ma et al. in 2002[2], where an aluminum ultra-thin layer was evaporated between two organic layers leading to the formation of nanoparticles into the organic material (2-amino-4,5-imidazoledicarbonitrile). These devices show reproducible switching with high ON/OFF current ratio of about $10^4$ and operation speed at the nanosecond.[2] Since this pioneering demonstration, many hybrid organic / metallic nanoparticles materials have been used to demonstrate resistive switching: polyaniline with gold nanoparticles[3], poly(3-hexylthiophene) (P3HT) with gold nanoparticles[4], Alq3 and various other small molecules with Au and Al nanoparticles[5], poly(N-vinylcarbazole) with silver sulfide nanoparticles[6] or with Au nanoparticles[7], polystyrene blended with Au nanoparticles capped with conjugated 2-naphthalenethiol[8] or with 1-dodecanethiol[9], a perylene derivative with Au nanoparticles[10], pentacene films with embedded Au nanoparticles[11;12]. Some of these previous reports claimed, in addition to resistive switching, the observation of a Negative Differential Resistance (NDR) in the device.[3;5;6] However in these previous works, devices have a vertical sandwich structure consisting in a metal/switching layer/metal stack where the hybrid material is localized between top and bottom electrodes. In these devices two approaches were used to form the hybrid material: (i) by mixing the organic material with the nanoparticles in solution and depositing the blend by spin-coating on the surface; or (ii) by evaporation of an ultra-thin metallic layer (usually 5 nm) onto the organic layer in order to form metal clusters or nanoparticles[13].

Furthermore, resistive memories and their extension to memristive systems have opened new routes toward innovative computing solutions such as logic-in-memory (implication logic with memristor)[14], analog computing (threshold logic with memristor[15]) or neuromorphic computing[16]. Nevertheless, to become attractive, these different approaches require a massive integration of memory elements. If crossbar integration consisting in vertical devices interconnected between metallic lines and columns has been considered as an interesting solutions, its practical realization with standard lithographic technics (i.e. top-down approaches) is facing severe limitations such as wire resistance contribution, crosstalk of memory elements during programming and sneak path during reading. One relatively unexplored solution is to rely on bottom-up approaches based on randomly assembly elements, in which memory functionalities are configured post-fabrication. This idea was initially proposed with the concept of nanocell[17;18;19] but remains in its early steps of development from a practical viewpoint and would benefit from functional and reliable hardware for its implementation.

Here, we report on a hybrid organic/gold nanoparticle memory with several advances compared to the devices demonstrated until now: (i) *a nanoscale –monolayer thick- planar structure*: the hybrid organic/gold nanoparticle material is placed between two coplanar electrodes with characteristic distance smaller than 100 nm; (ii) *a bottom-up approach for material synthesis, device fabrication and operation*: redox ligands were electro-polymerized in situ (i.e. into the device) to form a monolayer of a conjugated polymer with embedded Au nanoparticles and the memory functionality is defined post-fabrication to implement multi-terminals reconfigurable logic devices. This planar structure presents the main advantage, compare to vertical structure, to study the monolayer modifications during the voltage operations in order to understand the physical mechanisms involved during these operations. Also, this planar structure opens the way to multi-planar electrodes for connecting nanocell devices.

**METHODS**

**Electrode fabrication.** Devices were processed using a standard electron beam lithography process. We used highly-doped $n^+$-type silicon (resistivity 1-3 mΩ.cm) covered with a thermally grown 200 nm thick silicon dioxide (135 min at 1100°C in presence of oxygen 2L/min followed by a post-oxidation annealing at 900°C in N2 2L/min). The planar electrodes were patterned by electronic lithography with help of 10% MAA 17.5% / PMMA 3% 495K bilayer resists (with thicknesses of 510 nm and 85 nm respectively). Titanium/Platinum, (5/50 nm) were deposited by vacuum evaporation and lift-off. We fabricated electrodes with channel lengths L = 100 nm to 10 μm and channel width W = 100 nm to 1000 μm.

**Electrochemical experiments** were performed with a Modulab potentiostat from Solartron Analytical. The substrate with lithographed electrodes was hermetically fixed at the bottom of a 0.2 mL Teflon cell (see the setup photo in Supporting Information) containing the electrolyte solution. The counter electrode was a platinum wire (0.5 mm in diameter) and Ag/AgCl was used as a reference electrode.

**Electrical measurement setup.** For the Pt//PTEDOT-AuNPs//Pt switch test, electrical measurement were performed with an Agilent 4156C parameter analyzer in DC sweeping mode. We used Carl Süss PM5 probe station in order to connect Agilent 4156C and devices. For the pulse test, two programming and read pulses of square shape with different amplitude and width were applied to the device using the B1530A pulse generator module which is embedded in Agilent B1500A. All the electrical measurements were performed under ambient nitrogen in glove box filled with clean nitrogen ($O_2$ < 1 ppm, $H_2O$ < 1 ppm).

**Device fabrication.** *TEDOT-capped AuNPs*. The first step consists of the synthesis of 10 nm TEDOT-capped AuNPs. The synthesis of the ligand TEDOT-SH, consisting of a decylthiol chain terminated by a thienyl EDOT (TEDOT) end group was previously described[20]. The synthesis of 10 nm TEDOT-AuNPs (Fig. 1) involved a ligand exchange by treating 10 nm oleylamine-capped AuNPs[21] in the presence of TEDOT-SH. It was previously shown that oleylamine ligands are easily substituted by thiols[22;23;24] (see Supporting Information).

The ligand substitution is evidenced by UV-visible spectroscopy in 1,1,2,2-tetrachloroethane TCE solution (Supporting Information - Figure SI-2). The gold Surface Plasmon Resonance (SPR) of TEDOT-AuNPs is observed at 527 nm, while the absorption band corresponding to the TEDOT ligands is detected in the 315 - 353 nm region in agreement with previous results on the analogous 2-3 nm TEDOT-AuNPs directly synthesized with the ligand grafted on the Au NP by the Brust-Schiffrin method[20].

*TEDOT-AuNPs monolayer deposition*. In the second step, we formed the TEDOT-AuNPs monolayer on the device. For this, Langmuir films of TEDOT-AuNPs were prepared following the method of Santhanam[25] by evaporating a solution of functionalized NPs in TCE on the convex meniscus of a DI (deionized) water surface in a teflon petri dish (see details in Supporting Information, Figure SI-3). The transfer of the floating film on the substrate with lithographed electrodes was realized by dip coating. The TEDOT-AuNPs form a rather well organized monolayer on the surface (also called NPSAN: Nanoparticle Self-Assembled Network) as shown by scanning electron microscopy (SEM image, Fig. 2a). The statistical SEM image analysis gives an average diameter of the TEDOT-AuNPs of 9.3 nm, and an average spacing of 2.5 - 3.0 nm between the NPs in the network (see Supporting Information, Figure SI-4). Given the theoretical length of the free ligand (2.1 nm) obtained by a MOPAC simulation (ChemOffice Software) with respect to the average spacing measured between the NPs (2.5 to 3 nm), one can assume that the ligand molecules are interdigitated (Fig. 2b).

X-ray Photoelectron Spectroscopy (XPS) of the deposited TEDOT-AuNPs monolayer on large surface silicon without electrodes, shows the chemical composition of TEDOT adsorbates before electro-polymerization (results and spectra of XPS are presented in the Supporting Information-Figure SI-5). The good agreement between the theoretical and measured atomic ratios (see supporting information, table SI-1) demonstrates the successful grafting of TEDOT-SH on AuNPs. No signal is detected in the N1s region proving the complete substitution of oleylamine ligands by TEDOT-SH. Following the method of Volkert[26], we evaluate a ligand density of 4.0 molecules/nm$^2$ from the experimental S/Au ratio (0.407) of atomic concentrations.

*TEDOT-AuNPs monolayer electro-polymerization.* The third and last step for the device fabrication is the in-situ electro-polymerization of the TEDOT-AuNPs monolayer (last step in Fig 1). The monolayer of TEDOT-AuNPs deposited on and between Pt electrodes (see Fig. 2a) was electro-polymerized in situ, using the lithographed electrodes as working electrodes (see Materials and Methods) to form a monolayer film of conjugated polymer with embedded AuNPs, PTEDOT-AuNPs (see general principle in Fig. 3a). This polymerization of the monolayer was realized in potentiodynamic mode (electrolyte: 0.1M NBu$_4$PF$_6$ in CH$_2$Cl$_2$ or CH$_3$CN) by multiple scans at 100 mV/s between -0.4 V and +1 V. It leads to the development of a broad oxidation peak centered at +0.7 V vs Ag/AgCl which stabilized after multiple scans suggesting that most of the redox active TEDOT units have been coupled (Figure 3b). This behavior is clearly associated to the electro-polymerization of the PTEDOT polymer in the device.[20]. Then, the polymer was thoroughly rinsed with pure CH$_3$CN.

**RESULTS AND DISCUSSION**

**Forming process.** After the deposition and the in situ electro-polymerization, the PTEDOT is in its not conducting reduced form, as evidenced by the low current (70 nA at 5 V) measured for the current voltage (I-V) characteristic (Fig. 4). However, the current is higher than the one measured for the same device before in-situ electro-polymerization (around 1 nA at 5 V), because, in this former case, charge carriers have to tunnel between neighboring NPs though the TEDOT ligands.

A "forming process" is used to switch the PTEDOT-AuNPs monolayer in a more conducting state. For that, voltage sweeps in the range 0 V and 40 V (depending on the electrode spacing) at a sweep rate of around 4 V/s (we did not observed dependence of the sweep rate on the forming process), were rapidly applied, back and forth, on the device (Fig. 4). At the beginning, for the sweep voltage between 0 and 10 V (or a lateral electric field of 0.5 MV/cm, for a device length of 200 nm in that case) the forward and reverse curves were superimposed with a low noise (I-V curve labeled "unformed" in Fig. 4). Progressively, with the increase of the number of voltage sweeps and the increase of the maximum voltage (up to 20 V or 1 MV/cm), the I-V curves become more and more noisy. Finally, after several sweeps (three in the Fig. 4, and generally between 1 and 4), the current increases sharply at a voltage of about 6 – 7 V (curve labeled "formed" in Fig. 4), indicating the transition from a low conductivity to a higher conductivity state (the ON state). An instability region (at 6-7 V) systematically precedes this sharp increase with an important fluctuation on the measured current, and this voltage is independent of the electrode distance. This last point suggests the fact that the conduction is due to the formation of conducting paths between the electrodes after the forming process. The physical characterization by various technics of these conducting paths will be described in more details elsewhere[27]. In the "unformed" state the current is around 3.3 nA and 87 nA at 1 V and 5 V respectively. After forming, the device is in the ON state with currents at around 52 nA and 4.1 µA at 1 V and 5 V respectively, giving ratios of conductance of more than 2 orders of magnitude for the device shown in Fig. 4.

**NDR effect in the Formed-PTEDOT-AuNPs film.** After the electroforming process, the film is called Formed-PTEDOT-AuNPs thereafter, and the device remains in this ON state even after turning off the voltage power. Fig. 5 shows the I-V measured just after the forming process. The forward and backward current curves in the ON state are very close with high current values (around $10^{-4}$ A at 5 V).

Furthermore, the I-V characteristic after the electroforming process exhibits an interesting behavior: a NDR (negative differential resistance) with a maximum peak/valley ratio up to 17. This behavior is systematically and repeatedly observed for the different device geometries with a NDR peak in the range 5 - 7 V. For negative voltages, the I-V characteristics are identical with the presence of a NDR peak between -5 V and -7 V (see the Supporting Information – Figure SI-7).

**Control of the ON-to-OFF and OFF-to-ON switches in the Formed-PTEDOT-AuNPs film.**
As shown above, the device switches to the ON state during the forming process and remains stable in the ON state during a voltage sweep. One of the most important features of the devices is that an OFF state, with a current lower than in the unformed state, can be obtained if the voltage is very rapidly (<0.1s) returned to 0 V (Fig. 6). Fig 6 shows the I-Vs for a sequence of voltage sweeps as follow starting in the ON state : (1) 0 to 20 V at approximately 4 V/s, (2) 20 V to 0 V abruptly (< 0.1 s), (3) 0 to 20 V at approximately 4 V/s and (4) 20 V to 0 V at approximately 4 V/s. In the ON state and at low voltage sweep rates (curves 1 and 4) the device shows the NDR effect (also Fig. 5). During the abrupt sweep back to 0 V (trace 2, the current is not shown because it is not measurable during this abrupt event) the device switches in an OFF state as shown by the low current measured during the subsequent voltage sweep (trace 3 in Fig. 6) until a voltage of about 6 – 7 V where the current increases to recover the "valley" current of the NDR as measured during the voltage sweep (1). From these I-V curves, we can define two voltage domains: below 6-7 V for a memory effect with ON/OFF ratio in the range $10^3$-$10^4$ (see details in next section "Dynamic") and the NDR effect for the "ON" state at voltages > 6-7 V with a pic/valley ratio up to 17 (between 3 and 17 in for the devices measured in this work). The details of the dynamics of the ON/OFF switching is described and discussed in next section "Dynamic".

**Dynamic measurements in the Formed-PTEDOT-AuNPs film.** To determine the voltage amplitude and pulse width required to switch the device between the ON and OFF states, we designed the following experiments (Fig. 7a). For the ON to OFF switching (called RESET), the device is initially sets on the ON state by applying a double sweeping voltage (amplitude 15 V at 4 V/s) and the ON current $I_{ON}$ is measured by a small "reading" pulse (1 V and 50 ms). Then a writing pulse with voltage $V_{pulse}$ and duration $\Delta t$ was applied, again followed by a reading pulse to measure the current (Fig. 7a). The current ratio $I_{on}/I_{pulse}$ is a measurement of the ON-to-OFF ratio. These measurements were done for $V_{pulse}$ between 0 V and 16 V and $\Delta t$ from 1 μs to 1 s. The 3D plot in Fig. 7a shows that $V_{pulse} \gtrsim 8$ V is required to switch OFF the device device (i.e. $I_{ON}/I_{pulse} > 1$) and that a ON/OFF ratio of $10^3$-$10^4$ is obtained for voltage larger than about 10-12 V. This behavior is independent of $\Delta t$, or, in other words, it means that the switching time is lower than 1 µs. The transition from ON to OFF is voltage driven. On the contrary, the OFF-to-ON process (also called SET) is sensitive to both pulse amplitude and width (Fig. 7b). The same experiment was performed setting the device in the OFF state by applying a voltage sweep (15 V at 4 V/s) followed by an abrupt return to 0 V (as described previously in Fig. 6). In that case the current ratio $I_{pulse}/I_{off}$ measures the amplitude of the OFF-to-ON switching. Only pulses applied in a specific range: $5 V \lesssim V_{pulse} \lesssim 8$ V and with a pulse width $\Delta t_e \gtrsim 1$ ms, allow to switch ON the device.

**Data retention and endurance tests.** Endurance test was made to get further information on switching stability. Here, a "read" pulse (1 V during 50 ms) is added after each SET / RESET steps in order to achieve 'write-read-erase-read cycles' (Fig. 8a). According to results presented in Fig. 7, the switch from ON to OFF state is realized by appli a short (1 ms) voltage pulse of 12 V and the OFF to ON transition is induced by a medium voltage level of 6.5 V during 30 ms and a return to 0 V in 30 s (Fig 8a). As shown in figure 8a, these cycles are repeated for more than 800 cycles and display a current ratio $I_{on}/I_{off}$ of about $10^3$ without significant degradation. Data retention test shows the capability of the device to retain the data in the two resistance states. The device is switched to the ON or OFF states using the same voltage sequences as for the switching test (see Fig. 6), then we applied $10^5$ reading pulses at 1V (every 10 s) to measure the ON or OFF currents, respectively (Fig. 8b). The results (Fig. 8b) show a good data retention without significant current drift, revealing a non-volatile memory function of the device.

**NDR and memory mechanisms.** The NDR and memory behaviors observed here are similar as those described by Bozano et al.[28], in organic memory devices which were explained by the presence of metallic clusters in the organic films. Based on a review of literature results[29] and their own works, these authors established several criteria to describe this bistability in the device : (i) In the ON state, a "n" shaped I-V curve is observed with a maximum current at $V_{max}$ followed by a NDR effect (e.g. $V_{max} \sim$ 7 V in Fig. 6b); (ii) The OFF state is obtained by rapidly returning the voltage to 0V; (iii) The device remains in the OFF state until a threshold voltage $V_{th}$ is reached (e.g. $V_{th} \sim$ 6 V in Fig. 6b). Beyond $V_{th}$ the ON state is established; (iv) This threshold voltage is comparable to, but slightly less than $V_{max}$; (v) The ON state is obtained by setting the voltage above $V_{th}$, then reducing it to zero; (vi) Intermediate states, i.e., those between the ON and OFF resistance values, can be obtained by setting the voltage in the NDR region; (vii) Switching and reading are achieved with a single sign of applied voltage. All of these criteria established by Bozano et al. and observed by several authors (see for review[29] and[2-13]) are related to vertical macroscopic devices. Here, we demonstrate similar NDR and memory behaviors at a single monolayer level and for devices with nano-scale lateral dimensions.

The proposed physical mechanism is essentially the same as described by Simmons and Verderber on electroformed metal-insulator-metal diodes in the 1960s[30], involving charge trapping and space charge field inhibition of injection. Here, a model of charge trapping by NPs can also be proposed to explain this resistive switching behavior, considering the NPs as charge traps in the PTEDOT film. The transition speed between the ON and OFF states depends on how fast the charge carrier trapping by AuNPs takes place. Dynamics experiments (Fig. 7a) tell us that the trapping occurs with a characteristic time $\tau_C$ faster than 1 μs and requires a voltage larger than 8 V. These features are compatible with a field-assisted trapping mechanism characterized by a capture cross-section $\sigma_C$ larger than $(v_p p \tau_C)^{-1}$, with $v_p$ the thermal velocity of charge carriers (holes here), p the hole density and $\tau_C = 1 \mu s$. The OFF to ON transition corresponds to charge carrier emission by the NPs. Data in Fig. 7b shows an emission time $\tau_E > 1$ ms and that the applied voltage should be between 5 and 8 V. The emission time $\tau_E > 1$ ms implies that the energy level associated to the NP trap is located in the PTEDOT band gap, above the valence band, at an energy $E_T > kT Ln(\tau_E/\sigma_p v_p N_V)$ with $N_V$ the density of states in the valence band, k the Boltzmann constant, and T the temperature. The voltage dependent behavior, V > 5 V, may be explained by a field-assisted emission process (Poole-Frenkel). The upper limit, V < 8 V, can correspond to a competition between capture and emission, this voltage being similar to the one observed for the capture process (see details in the Supporting Information). To go further, more experiments are required to determine the parameters involved in the SHR (Shockley-Hall-Read) equations ($v_p$, p, $N_V$), as well as temperature dependent measurements to determine the energy level $E_T$. Moreover, this charge trapping/detrapping model in organic memory is also the subject of discussion in the literature[28;31], and a more detailed discussion is beyond the scope of this paper. More results on the ON/OFF switching in these monolayer devices as well on the forming process will be reported elsewhere[27].

**Reconfigurable logic functions.** These memory features can be conveniently integrated into multi-terminal devices by following a post-fabrication approach. First, the device location between different coplanar multi-electrodes (here 6, Fig. 9a) spaced by 100 nm is defined by following the forming process described above. In figure 9a, two inputs (IN1 and IN2) with a common output (OUT) are randomly selected out of the 6 terminals to define two reconfigurable memory elements (i.e. located between IN1/OUT and IN2/OUT, respectively) and the previously described forming process is sequentially applied on these two devices. To demonstrate the reconfigurability of the resulting multi-terminal devices, we choose to implement a simple reconfigurable logic function with two inputs and one output. This simple system is equivalent to the nanocell device proposed in [17;18;19] and corresponds to an important step toward the material implementation of this kind of device. Figure 9 presents the reconfiguration of the multi-terminal device to AND and OR logic functions. The two devices are programmed by sweeping the voltage at 4V/s between the two electrodes (between IN1/OUT and IN2/OUT in Fig. 9a) according to the rules shown in Fig. 6 : (i) 0 to 12 V and 12 to 0 V at 4 V/s for the ON state, (ii) 0 to 12 V and 12 V to 0 V abruptly (<0.1 s) for the OFF state or (iii) 0 to 10 V and 10 V to 0 V abruptly (< 0.1 s) for the intermediate resistance. Thus, the two devices (IN1/OUT and IN2/OUT) can independently be programmed to three different resistances. Reading of the nanocell state is realized by applying simultaneously to the two input terminals square shape pulses of 1 V and 5 V for implementing the logical "0" and "1", respectively (Fig. 9b). Fig. 9c shows the resulting output current when the two devices are programmed in the ON state resulting in a Boolean OR logic operation, (with respect to an arbitrary constant threshold current of 60 nA). By reconfiguring the memory element to an intermediate resistance state (case iii above), the multi-terminal device can be configured to a logic AND function (Fig. 9d). Programming the two devices in their OFF state results in no logic operation (i.e. the output current is always below the threshold value of 60 nA). Obviously, the random choice of the input /output terminals represents a critical aspect of our devices. Out of 18 equivalent 2 inputs/1 output combinations characterized, this reconfigurable Boolean logic functions is observed 55 %. This yield can be ascribed to variabilities in both the deposition, the in-situ electro-polymerization of the molecules/NPs networks and variabilities in the forming process. Future work will investigate in more detail the dependency between overlapping conducting paths and the realization of larger multi-terminal devices.

**CONCLUSION**

In summary, we demonstrated organic memristive devices realized by a monolayer-thick 2D network of gold nanoparticles functionalized by EDOT-thiophene molecules deposited between two planar electrodes with nanometer gap. The network is electro-polymerized in situ to form a 2D monolayer film of polymer PTEDOT with embedded AuNPs. After a forming process, the devices show bistable memory behavior with ON/OFF current ratios in the range $10^3$ - $10^4$. A NDR behavior with a peak/valley ratio up to 17 is systematically observed when the devices are in the ON state. The dynamic of the switching and programming voltage window are investigated. The observed switching behavior is compatible with a model of charge trapping/detrapping by the nanoparticle. A data retention test over more than 1 day and endurance test over 800 switching cycles are demonstrated. Finally, reconfigurable Boolean logic functions, with multi-terminal device inspired by the concept of nanocell are implemented.

## ASSOCIATED CONTENT

**Supporting Information**. Synthesis of TEDOT-AuNPs, preparation of TEDOT-AuNPs monolayers, XPS results, electro-polymerization of the TEDOT-AuNPs monolayers, NDR in both bias polarities and trapping/detrapping mechanisms. "This material is available free of charge via the Internet at http://pubs.acs.org."

## AUTHOR INFORMATION

**Corresponding Author**

* E-mail: stephane.lenfant@iemn.univ-lille1.fr.

## ACKNOWLEDGMENT

This work has been financially supported by EU FET project n° 318597 "SYMONE", by the ANR agency, project n° ANR 12 BS03 010 01 "SYNAPTOR" and the French RENATECH network.

The authors thank the clean room staff of the IEMN for the assistance and help for the device fabrication.

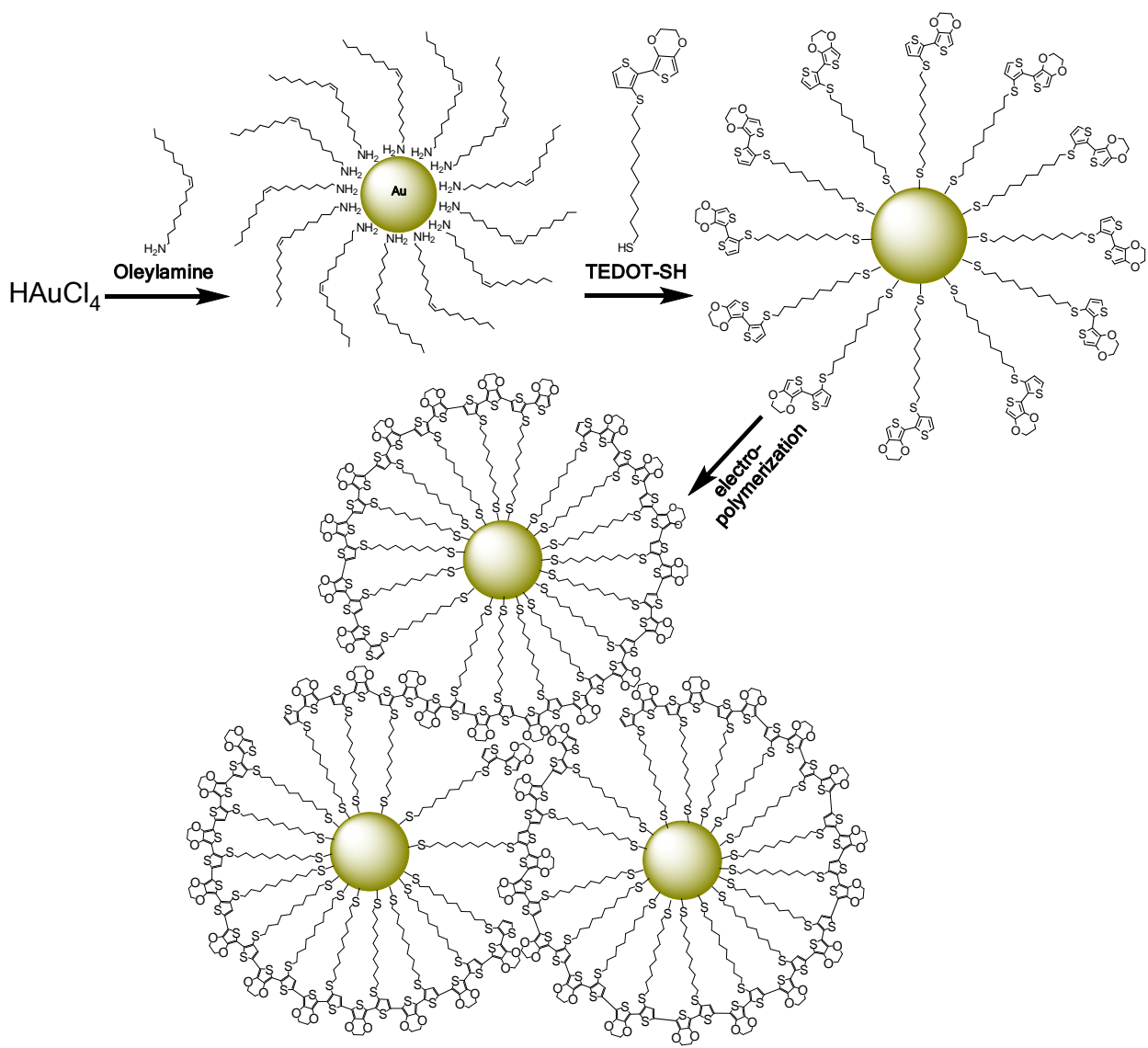

**Figure 1.** Schematic of the synthesis of the 10 nm PTEDOT-AuNPs in three steps: first synthesis of oleylamine capped 10 nm AuNPs from HAuCl$_4$, then exchange of the ligand with TEDOT-SH molecules and electro-polymerization of the TEDOT-AuNPs.

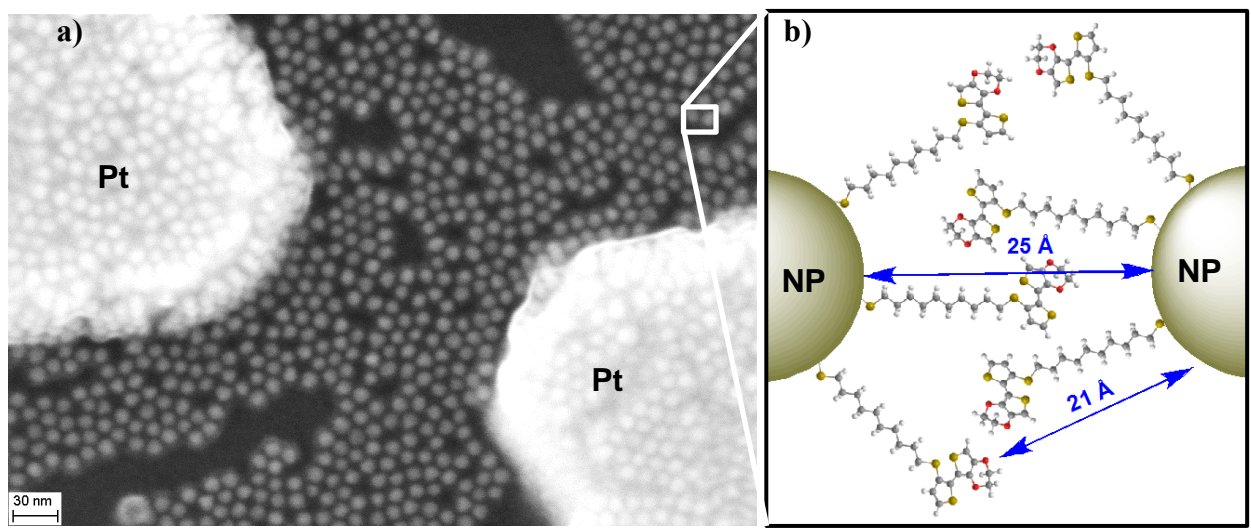

**Figure 2.** (a) SEM image showing the TEDOT-AuNPs monolayer deposited on a nanogap platinum electrodes (electrode spacing L = 200 nm and electrode width W=100 nm); bar scale represents 30 nm; (b) Magnified view of an NP interspace.

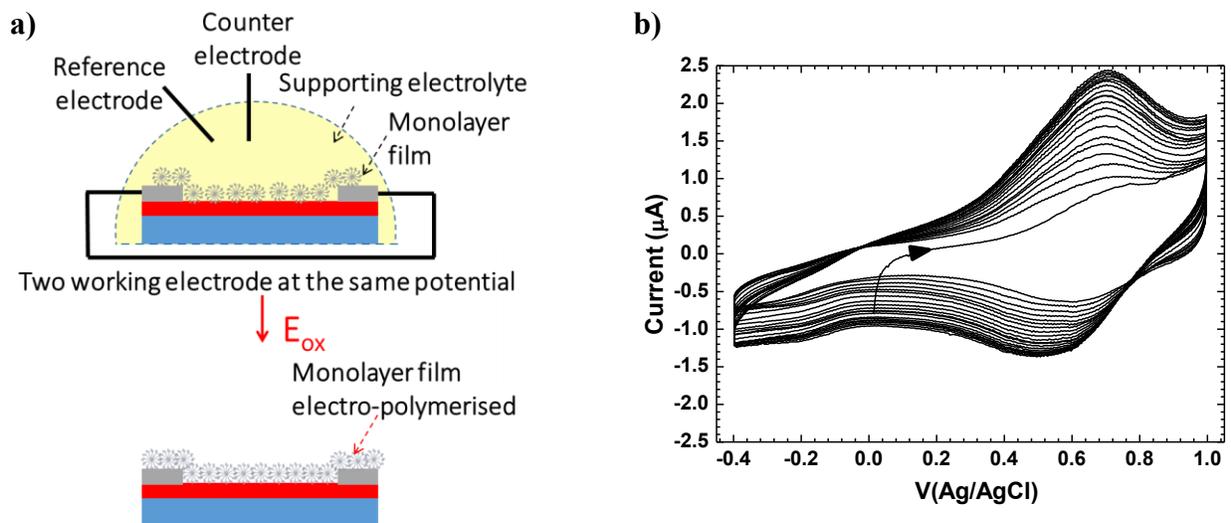

**Figure 3.** (a) Scheme of the in-situ electro-polymerization to form the monolayer between the electrodes; (b) Electro-polymerization of the TEDOT-AuNPs Langmuir film into a polymer PTEDOT with embedded AuNPs (PTEDOT-AuNPs) realized in potentiodynamic mode (electrolyte: 0.1M NBu$_4$PF$_6$ in CH$_2$Cl$_2$ or CH$_3$CN) by multiple scans at 100mV/s.

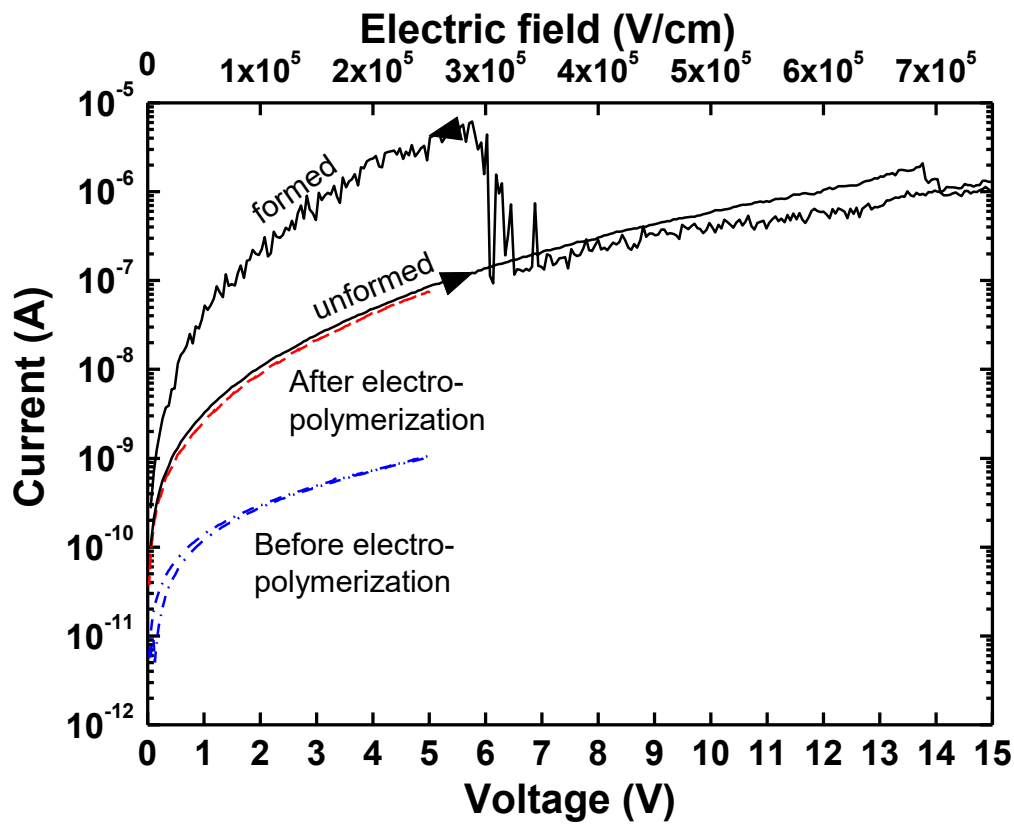

**Figure 4.** Typical I-V characteristic measured before electro-polymerization (dash-dot blue curve), after electro-polymerization (dash red curve) and during the forming process (black curve) on PTEDOT-AuNPs films device with a length and width of 200 nm and 100 nm, respectively for the electrode gap. Voltage sweep rate around 4 V/s.

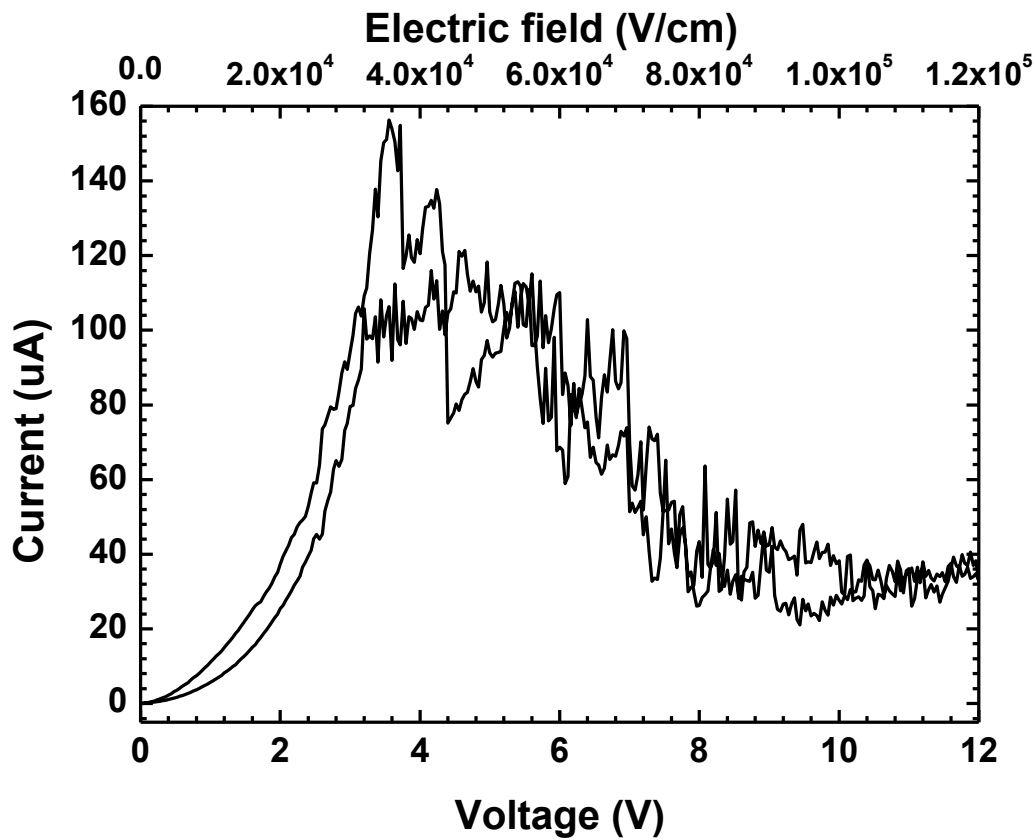

**Figure 5.** Current-voltage curves during a back and forth voltage sweeps in the ON state, showing NDR effect, and measured after the forming process on Formed-PTEDOT-AuNPs device with a length L and width W of 1 μm and 1 mm respectively, for the electrode gap. Voltage sweep rate around 4 V/s.

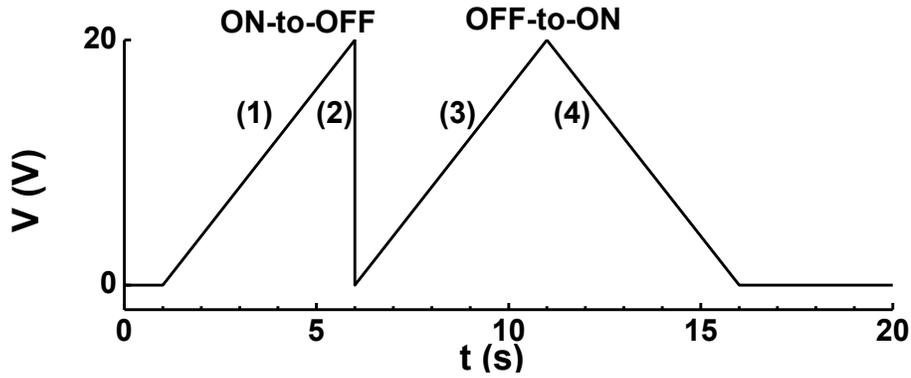

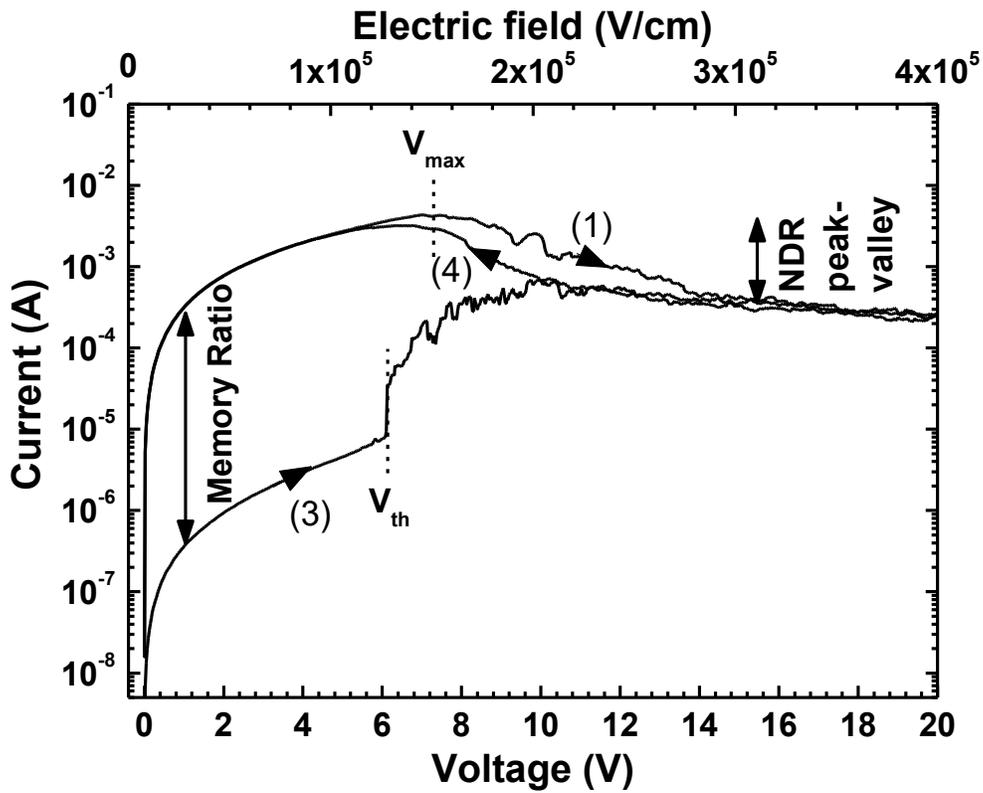

**Figure 6.** (a) Typical voltage sequence for the memory behavior and (b) the corresponding measured I-V curves for the ON-to-OFF and OFF-to-ON switches on a Formed-PTEDOT-AuNPs device with a length L = 500 nm and width W = 1 mm for the electrode gap. The I-V traces are numbered in accordance with the voltage sequences shown in (a).

a)

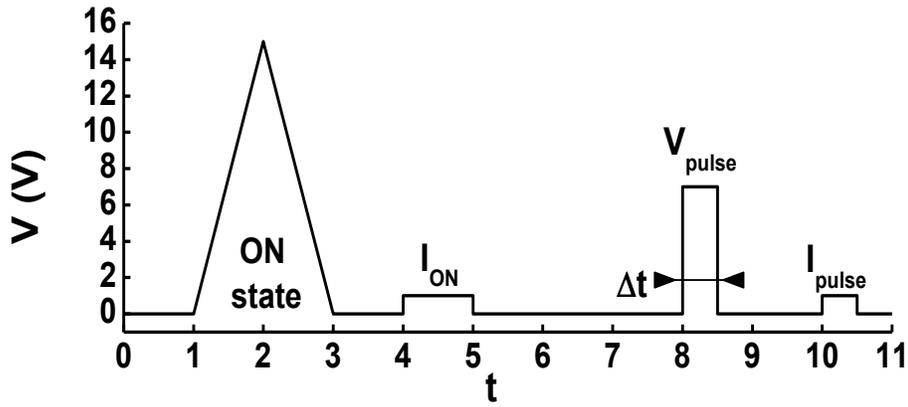

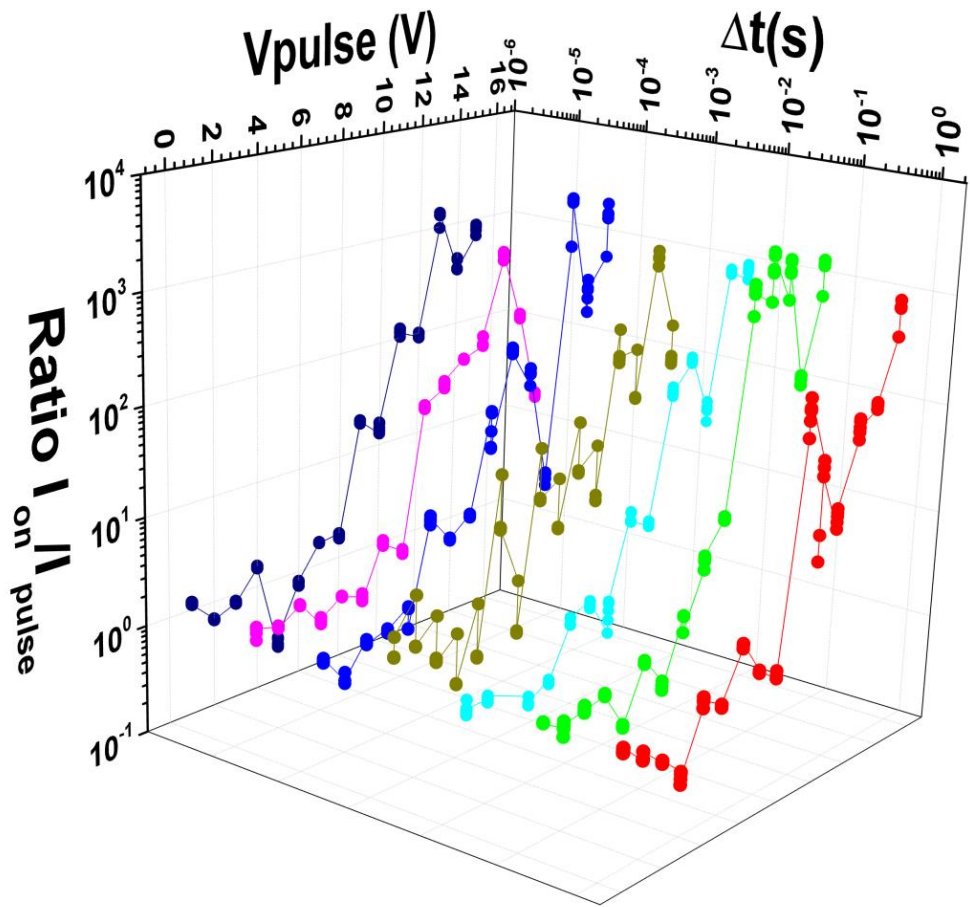

b)

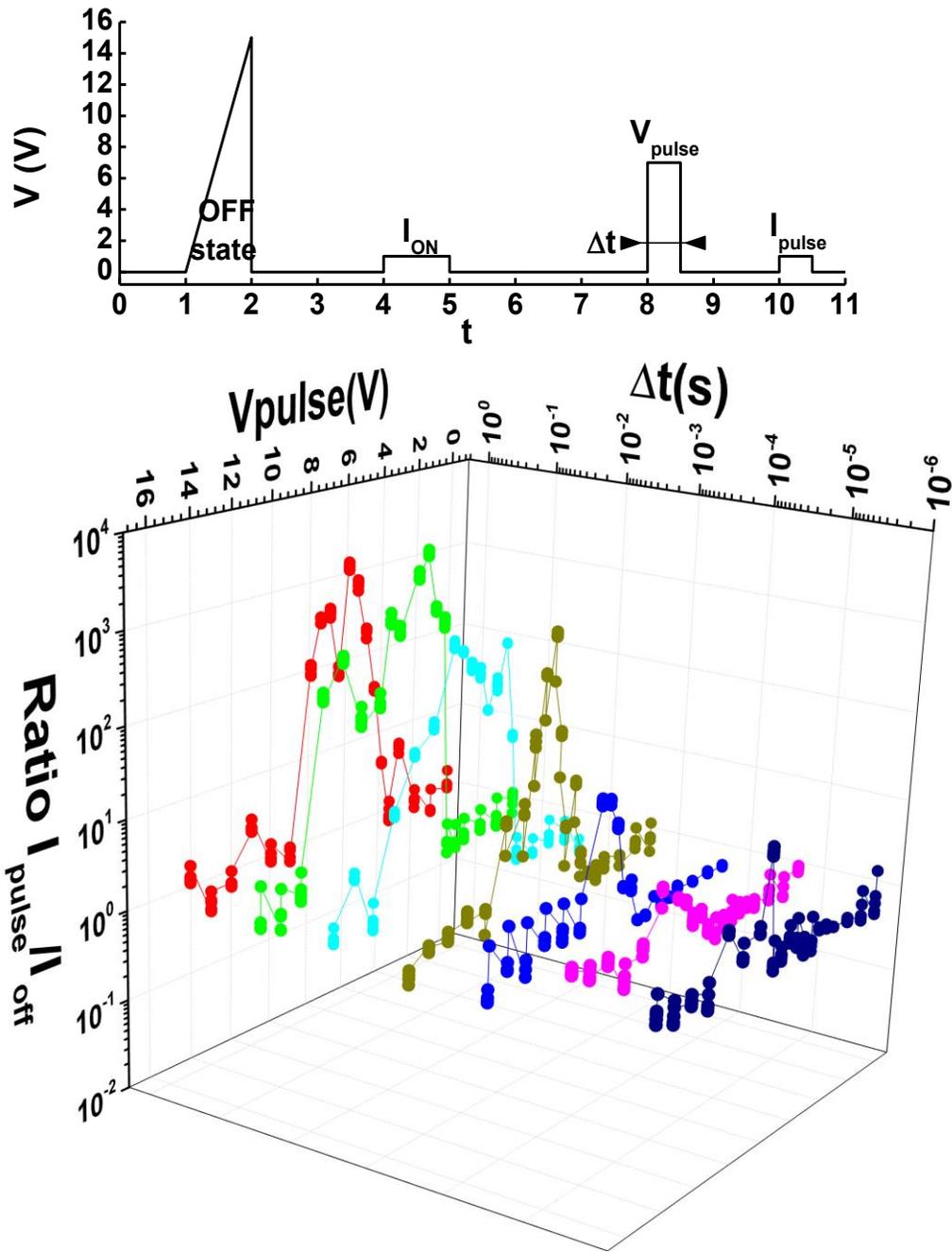

**Figure 7.** Dynamic measurements on Formed-PTEDOT-AuNPs devices with length and width of 500 μm and 1 mm, respectively: (a) ON-to-OFF dynamics and voltage windows. The switching is time independent in the range 1 μs - 1 s, and it requires a voltage ≳ 8 V; (b) OFF-to-ON dynamics and voltage windows. The switching requires a pulse duration > 1 ms and a voltage window between 5 and 8 V.

a)

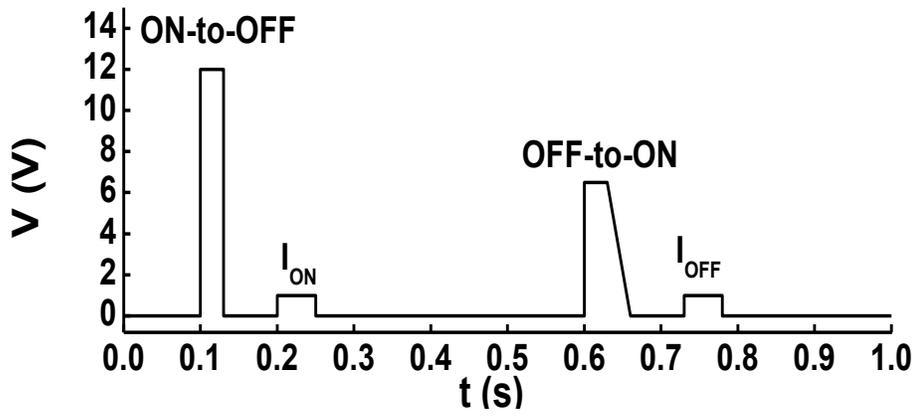

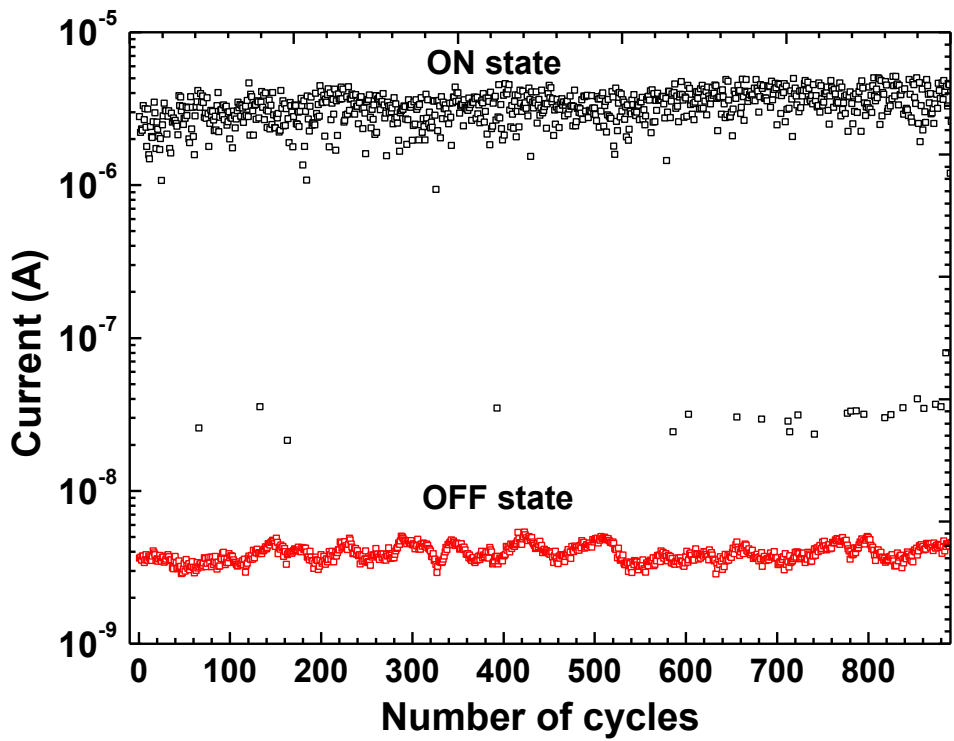

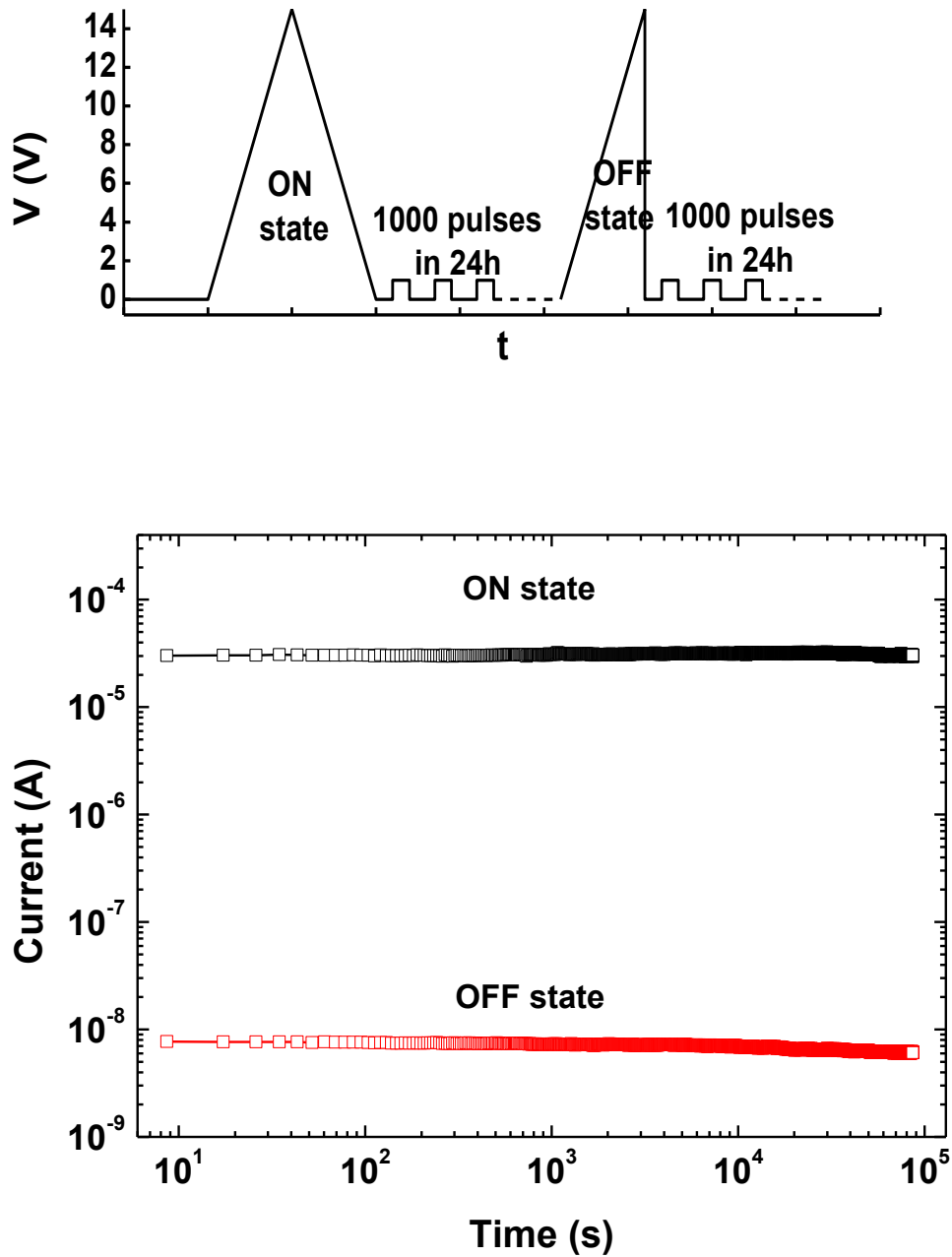

**Figure 8.** (a) Cycling experiments between ON and OFF states; and (b) data retention experiments showing the evolution of the ON state and OFF state currents.

a)

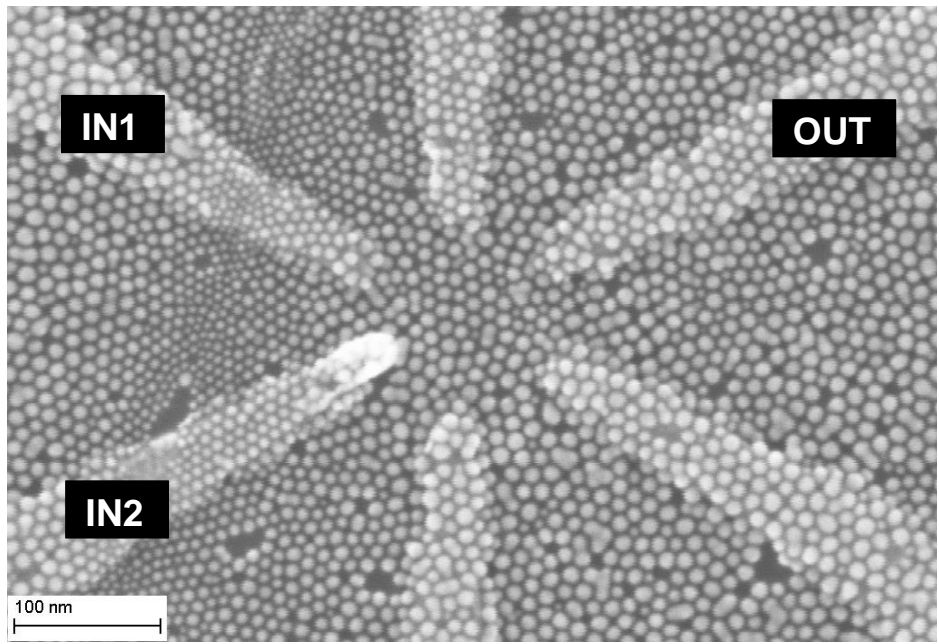

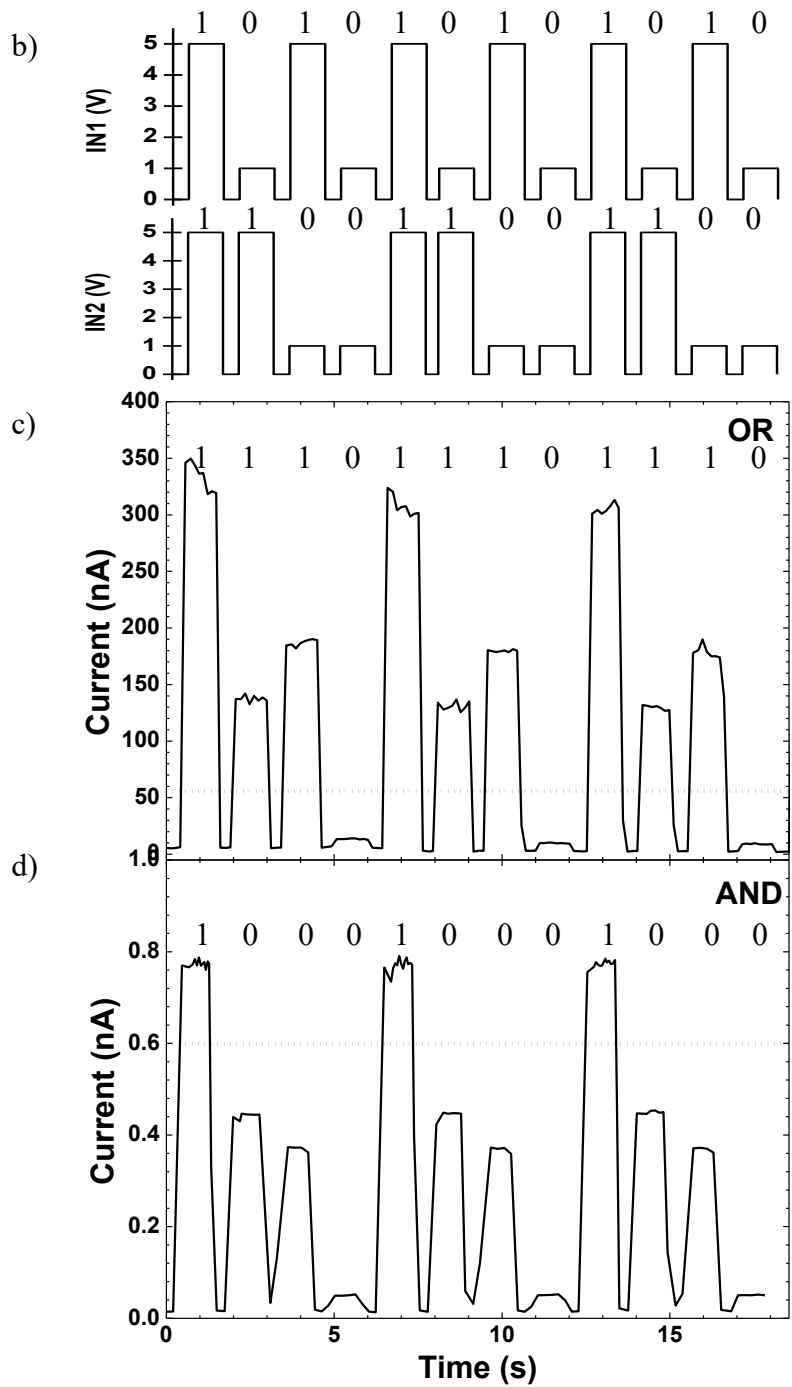

**Figure 9.** (a) SEM image showing the coplanar multi-electrodes spaced by 100 nm, two inputs (named IN1 and IN2) and 1 output (named OUT) are chosen on order to define two memory elements (i.e. located between IN1/OUT and IN2/OUT); bar scale represents 100 nm; (b) the pulse

stream applied on IN1 and IN2; (c) current measured at the OUT electrode after programming the two devices in the ON state; (d) current measured at the OUT electrode after programming the two devices in intermediate resistance state.

**TOC GRAPHIC**

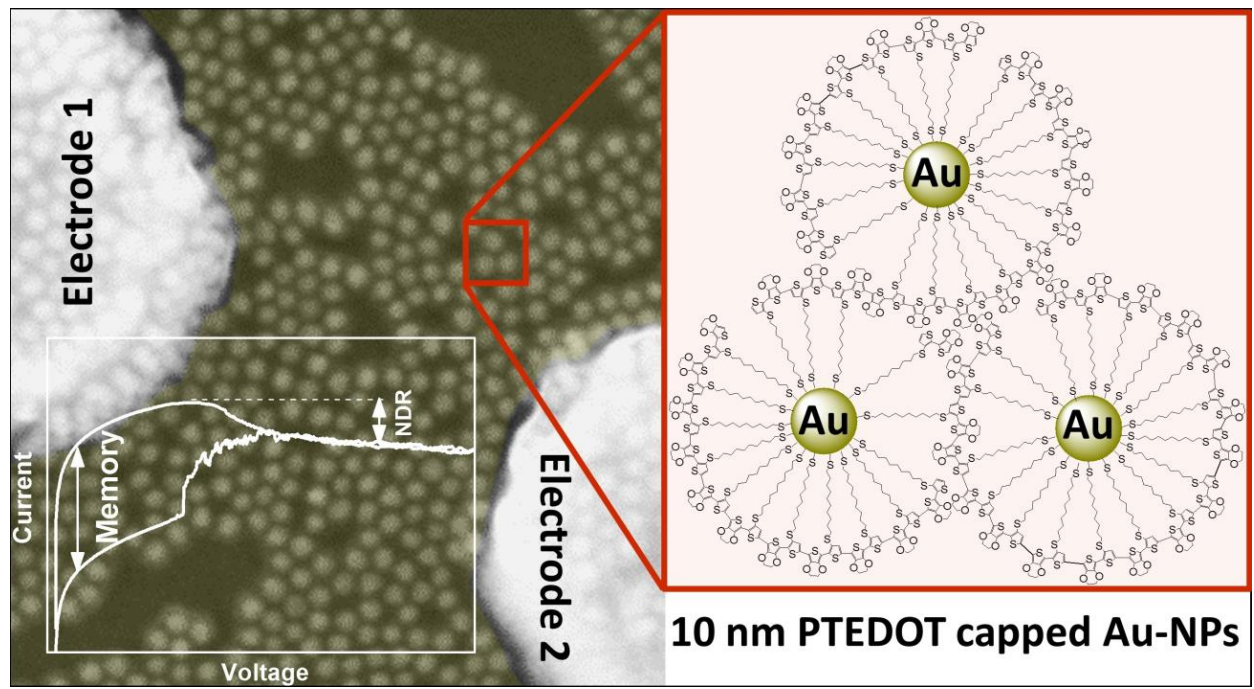

# Negative Differential Resistance, Memory and Reconfigurable Logic Functions based on Monolayer Devices derived from Gold Nanoparticles Functionalized with Electro-polymerizable Thiophene-EDOT Units

T. Zhang[1], D. Guérin[1], F. Alibart[1], D. Vuillaume[1], K. Lmimouni[1], S. Lenfant[1*]

A. Yassin[2], M. Oçafrain[2], P. Blanchard[2], J. Roncali[2]

[1] Institute of Electronics Microelectronics and Nanotechnology (IEMN), CNRS, University of Lille, Avenue Poincaré, 59652 Villeneuve d'Ascq, France

[2] MOLTECH-Anjou, CNRS, University of Angers, 2 Bd. Lavoisier, Angers, 49045, France

## SUPPORTING INFORMATION

# 1. SYNTHESIS OF TEDOT-AuNPs

First, oleylamine capped AuNPs were synthesized using the Santhanam's procedure [1] 50 mg of $HAuCl_4.3H_2O$ in 5 ml of oleylamine (95%) and 5 ml of anhydrous toluene were dissolved in a schlenk flask equipped with a nitrogen inlet and a condenser. The homogenous orange solution was maintained under nitrogen atmosphere at 80°C under stirring for 10h. During this time, the color changed from orange to colorless then to dark red. The NPs purification was performed by repeated centrifugation/sonication cycles of the solution: the reaction mixture was first diluted with 30 % of hexane, followed by addition of ethanol to precipitate the NPs. After centrifugation for 5 min at 7000 tr/min, the supernatant was eliminated, then the NPs were redispersed in hexane by sonication. This process was repeated three times and finally the oleylamine-AuNPs were redissolved in 5 mL of toluene (good stability in this solvent several months at 5°C). By the same precipitation/centrifugation process, toluene was easily replaced by dimethylsulfoxide (DMSO) for the ligand exchange reaction with TEDOT-SH (see below). The surface plasmon resonance peak (SPR) is observed at 524 nm in $CHCl_3$ by UV-vis spectroscopy. Average diameter of NPs measured by the statistical analysis of SEM image is 9.9 nm (Figures SI-1b and c).

---

[1] Santhanam, V.; Liu, J.; Agarwal, R.; Andres, R. P., Self-assembly of uniform monolayer arrays of nanoparticles. Langmuir 2003, 19 (19), 7881-7887.

a)
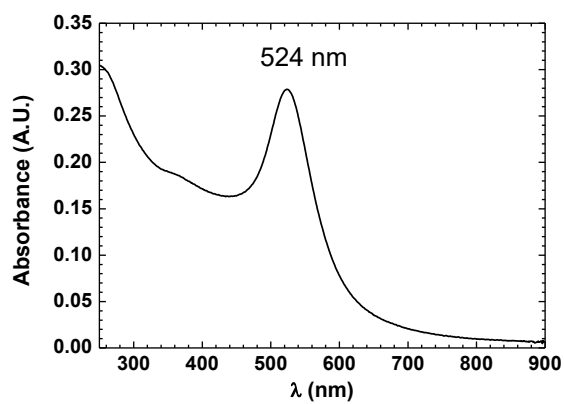

b)
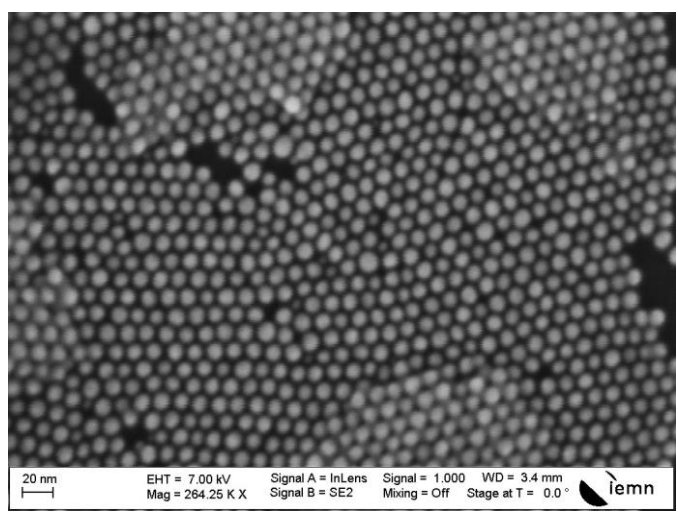

c)
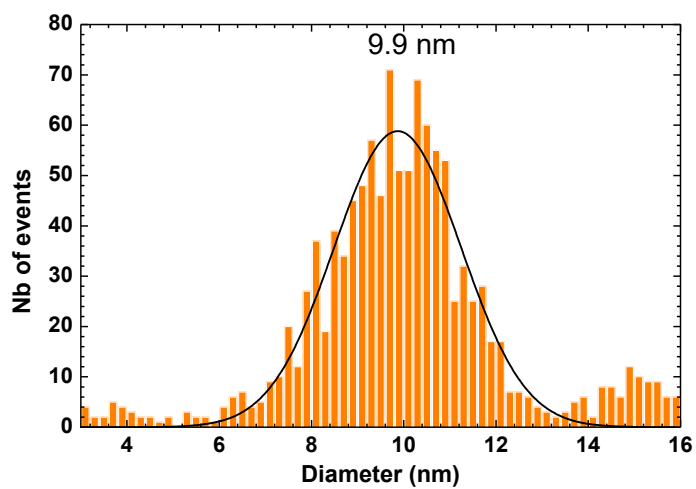

**Figure SI-1**. a) UV-vis spectrum of oleylamine-AuNPs in $CHCl_3$; (b) SEM image of the oleylamine-AuNPs monolayer deposited on flat $SiO_2$ surface; c) and the corresponding average diameter statistical image study.

Secondly, the synthesis of TEDOT-AuNPs was carried out from oleylamine capped AuNPs. Dimethylsulfoxide (DMSO) turned out appropriate solvent for EDOT-AuNPs synthesis whereas water-immiscible 1,1,2,2-tetrachloroethane (TCE) was a suitable solvent for the preparation of Langmuir films. Solvents were perfectly degassed by nitrogen bubbling. In a nitrogen glovebox ($O_2$ and $H_2O$ < 5 ppm), 5 mg of TEDOT-SH (synthesis described elsewhere [2]) were added to 1 mL of the previous oleylamine-AuNPs solution in DMSO. The thiolation was performed at 60°C for 6h under nitrogen in the dark. The ligand excess was eliminated as following: the reaction mixture was first diluted with 50 % of toluene (solvents in which TEDOT-AuNPs are few soluble) then centrifuged for 3 min at 7000 tr/min. The supernatant was eliminated then the precipitate was cleaned thoroughly by fresh toluene. Finally, a concentrated red-purple solution of TEDOT-AuNPs was obtained by sonication in 1 mL of TCE. We observed a good stability of this solution over long period at 5°C (several weeks). TEDOT-AuNPs average diameter measured by SEM on a surface after film formation (see below) is 9.3 nm. The reproducibility of the synthesis of the Au-NPs with this approach is quite good, with a low dispersion of the average diameter of the NPs (< 1 nm). The ligand substitution is evidenced by UV-visible spectroscopy in TCE solution (Figure SI-2). The gold Surface Plasmon Resonance (SPR) of TEDOT-AuNPs is observed at 527 nm, while the absorption peaks corresponding to the TEDOT ligands is detected in the 315 - 353 nm region in agreement with previous results on the analogous 2 nm TEDOT-AuNPs directly synthesized with the ligand grafted on the NP by the Brust-Schiffrin method [3].

---

[2] Yassin, A.; Ocafrain, M.; Blanchard, P.; Mallet, R.; Roncali, J., Synthesis of Hybrid Electroactive Materials by Low-Potential Electropolymerization of Gold Nanoparticles Capped with Tailored EDOT-Thiophene Precursor Units. Chemelectrochem 2014, 1 (8), 1312-1318.

[3] Hiramatsu, H.; Osterloh, F. E., A simple large-scale synthesis of nearly monodisperse gold and silver nanoparticles with adjustable sizes and with exchangeable surfactants. Chemistry of Materials 2004, 16 (13), 2509-2511.

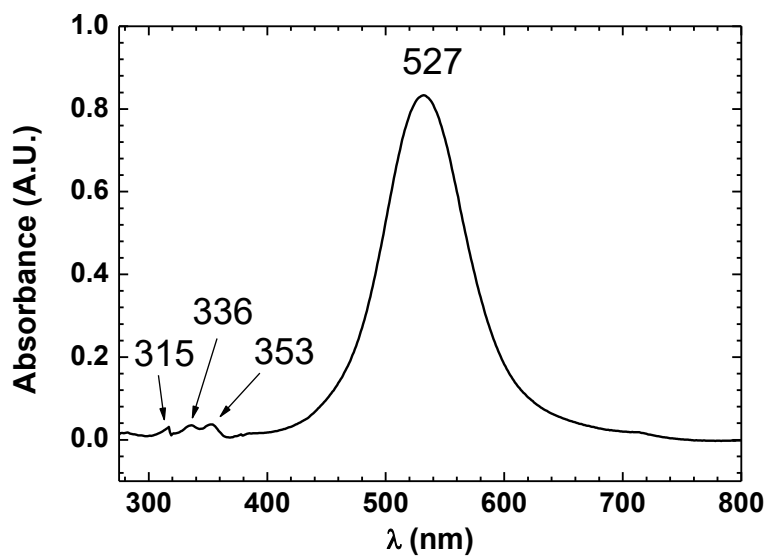

**Figure SI-2.** UV-vis spectrum of TEDOT-AuNPs in TCE (with base line correction).

## 2. PREPARATION OF THE TEDOT-AuNPs NPSAN MONOLAYER.

The preparation of NPSAN (Nanoparticles self-assembled networks) was realized following the method of Santhanam [1]. 100 µL of the previous EDOT-AuNPs solution in TCE were spread on a convex water meniscus delimited in a pierced teflon petri dish (hole diameter: 2 cm, see Figure SI-3). Once the half of TCE is evaporated, we observed that the addition of hexane (~50 µL) favored the formation of a compact NPSAN. After complete evaporation of TCE and hexane, the floating film was deposited on a clean substrate (lithographed substrate or silicon dioxide surface without electrodes). The transfer of the floating film was subsequently realized by dip coating directly on a lithographed substrate. After drying in the air, the film was rinsed with ethanol.

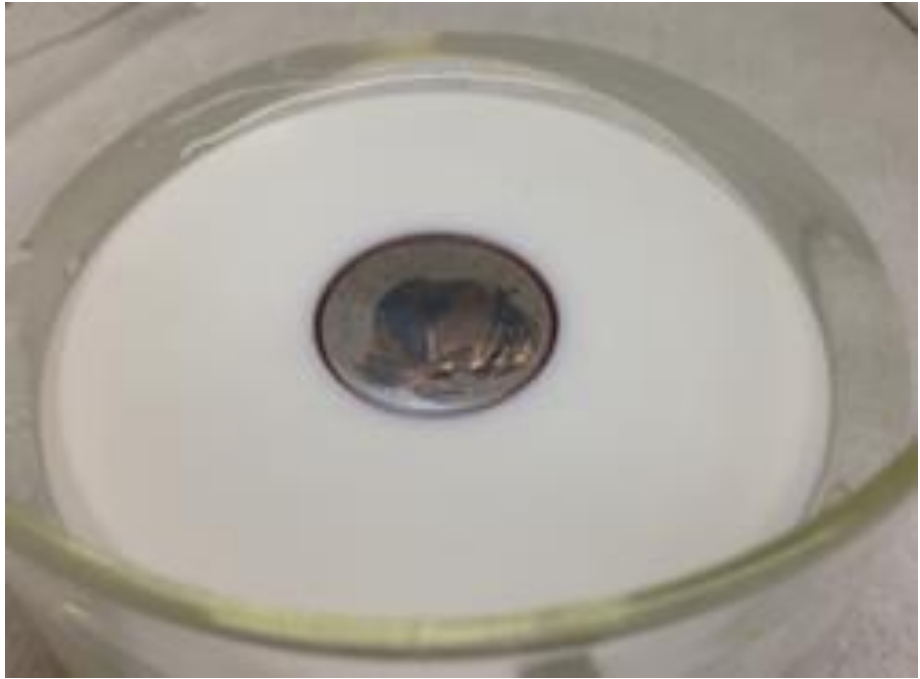

**Figure SI-3.** Monolayer film floating on the water surface in a Teflon dish.

The statistical analysis of the NP size was measured by SEM when the NPSAN is transferred onto the lithographed substrate (see Fig. 2a, main text), the average diameter of the TEDOT-AuNPs is 9.3 nm with a medium spacing of 2.5 - 3.0 nm within the network (see diameter dispersion in figure SI-4).

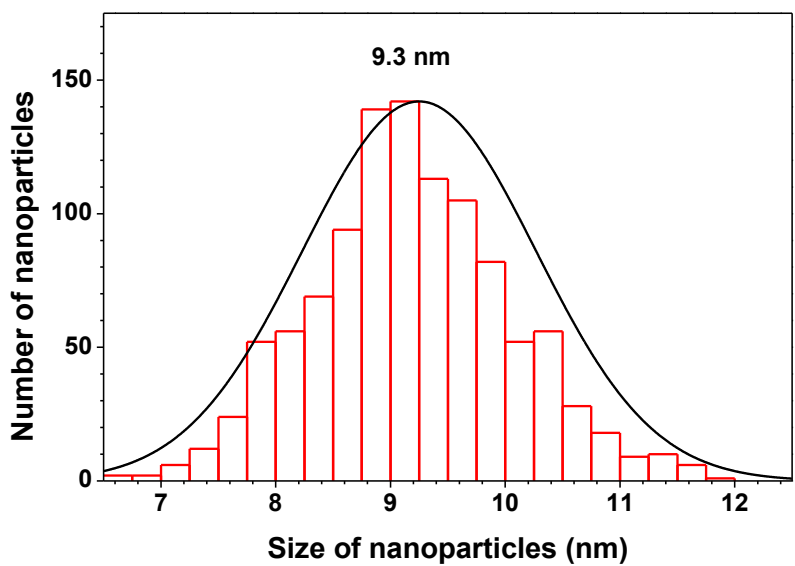

**Figure SI-4:** Statistical analysis of the TEDOT-AuNPs diameter: average value of 9.3 nm and FWHM of 1.2 nm. Average spacing between the NPs: 2.5 - 3 nm.

## 3. XPS SPECTRA ON TEDOT-AuNPs MONOLAYER.

X-ray Photoelectron Spectroscopy was carried out on the deposited TEDOT-AuNPs monolayer on large silicon dioxide surface without electrodes, to study the chemical composition of TEDOT adsorbates before electro-polymerization (results of XPS summarized in Table SI-1, and XPS spectra in Fig. SI-5). The C1s signal show two peaks corresponding to the different environments of carbon atoms in the EDOT ligand (C-C and C-S at 284.8 eV and C-O at 286.5 eV) with an experimental ratio C-O/(C-C+C-S) = 0.27 close to expected value (0.25). Likewise, ratio of experimental atomic concentration $S_{total}/C_{total}$ (0.22) was in accordance with expected value (4/20 = 0.20). The S2p signal is deconvoluted into a pair of doublets with a 1.2 eV splitting energy centered at 164.5 and 162.6 eV ascribed to sulfur bound to carbon (thiophene) and sulfur bound to gold, respectively. The characteristic binding energy of the $S2p_{2/3}$ at 162.0 eV is in agreement with what is generally found for organothiols chemisorbed on Au. [4;5;6] Analysis of the atomic ratios demonstrate that the grafting of TEDOT-SH on AuNPs was successful. No signal is detected in the N1s region proving the quantitative substitution of oleylamine ligands by TEDOT-SH. Following the method of Volkert [6], we evaluate a ligand density of 4.0 molecules/nm$^2$ from the experimental S/Au ratio (0.407) of atomic concentrations.

| *Attribution* | *Binding Energy (eV)* | *Area (A.U.)* | *FWHM (eV)* | *Atomic ratios* |
|---|---|---|---|---|
| *C1s-C+C1s-S* <br> *C1s-O* | 284.86 <br> 286.54 | 17067 <br> 4577 | 1.29 <br> 1.32 | C-O/(C-C+C-S) = 0.27 (0.25) |
| *S2p$_{3/2}$-Au* <br> *S2p$_{1/2}$-Au* <br> *S2p$_{3/2}$-C* <br> *S2p$_{1/2}$-C* | 162.02 <br> 163.15 <br> 163.90 <br> 165.22 | 679 <br> 338 <br> 2279 <br> 1095 | 1.34 <br> 1.34 <br> 1.34 <br> 1.34 | S$_{total}$/C$_{total}$ = 0.22 (0.20) <br> S-C/S-Au = 3.30 (3.00) <br> S$_{total}$/Au$_{total}$ = 0.407 |
| *Au4f$_{total}$* | 84.00, 87.70 | 10783 | | |

**Table SI-1.** XPS analysis of TEDOT-AuNPs NPSAN deposited on large silicon dioxide surface without electrodes. Areas are corrected by relative sensitivity factors [7]. Experimental values of atomic concentration ratios are compared to theoretical values (in brackets).

Data from Table 1 are calculated from the following XPS spectra of TEDOT-AuNPs monolayer.

a)

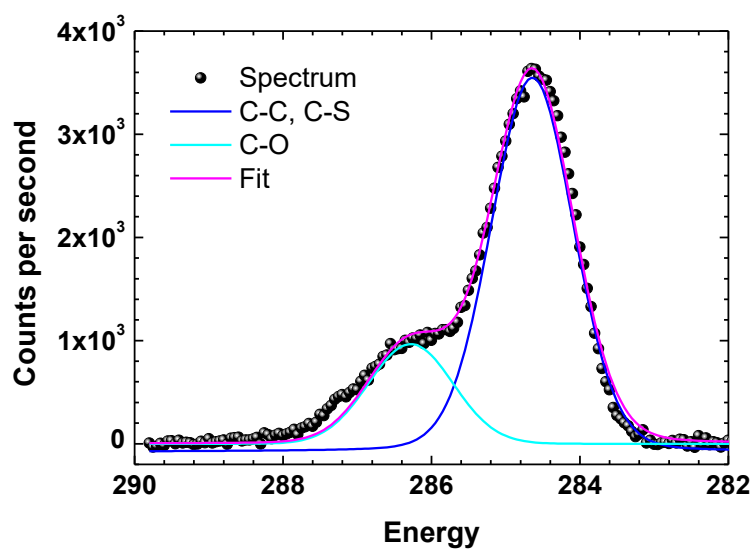

b)

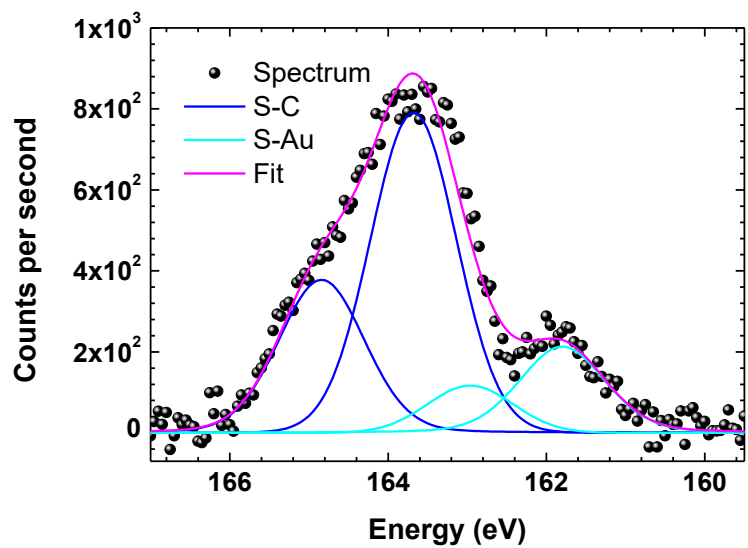

**Figures SI-5**. XPS spectra of TEDOT-AuNPs monolayer before electro-polymerization in the range of (a) C1s and (b) S2p regions with the peaks deconvolution.

## 4. ELECTRO-POLYMERIZATION OF THE TEDOT-AuNP MONOLAYER.

Electrochemical experiments were performed with a Modulab potentiostat from Solartron Analytical. Platinum working electrodes (WE) were lithographed on a silicon substrate covered with 200 nm $SiO_2$ oxide with channel lengths comprised between 100 nm and 10 µm (as described in the Methods section in the main text). This substrate was hermetically fixed at the bottom of a 0.2 mL Teflon cell (see Figures SI-6a and b) containing the electrolyte solution. The counter electrode (CE) was a platinum wire (0.5 mm) and Ag/AgCl was used as a reference electrode (RE) (see Figure SI-6a). The monolayer of TEDOT-AuNPs deposited on Pt working electrodes was electro-polymerized in situ to form a monolayer film of polymer with embedded AuNPs (PTEDOT-AuNPs) (see Figure SI-6c and Fig. SI-6d for the reaction). This polymerization of the monolayer was realized in potentiodynamic mode (electrolyte: 0.1M $NBu_4PF_6$ in $CH_2Cl_2$ or $CH_3CN$). The short channel length promotes the electro-polymerization of the entire TEDOT-AuNPs material localized between the two platinum working electrodes during the electro-polymerization.

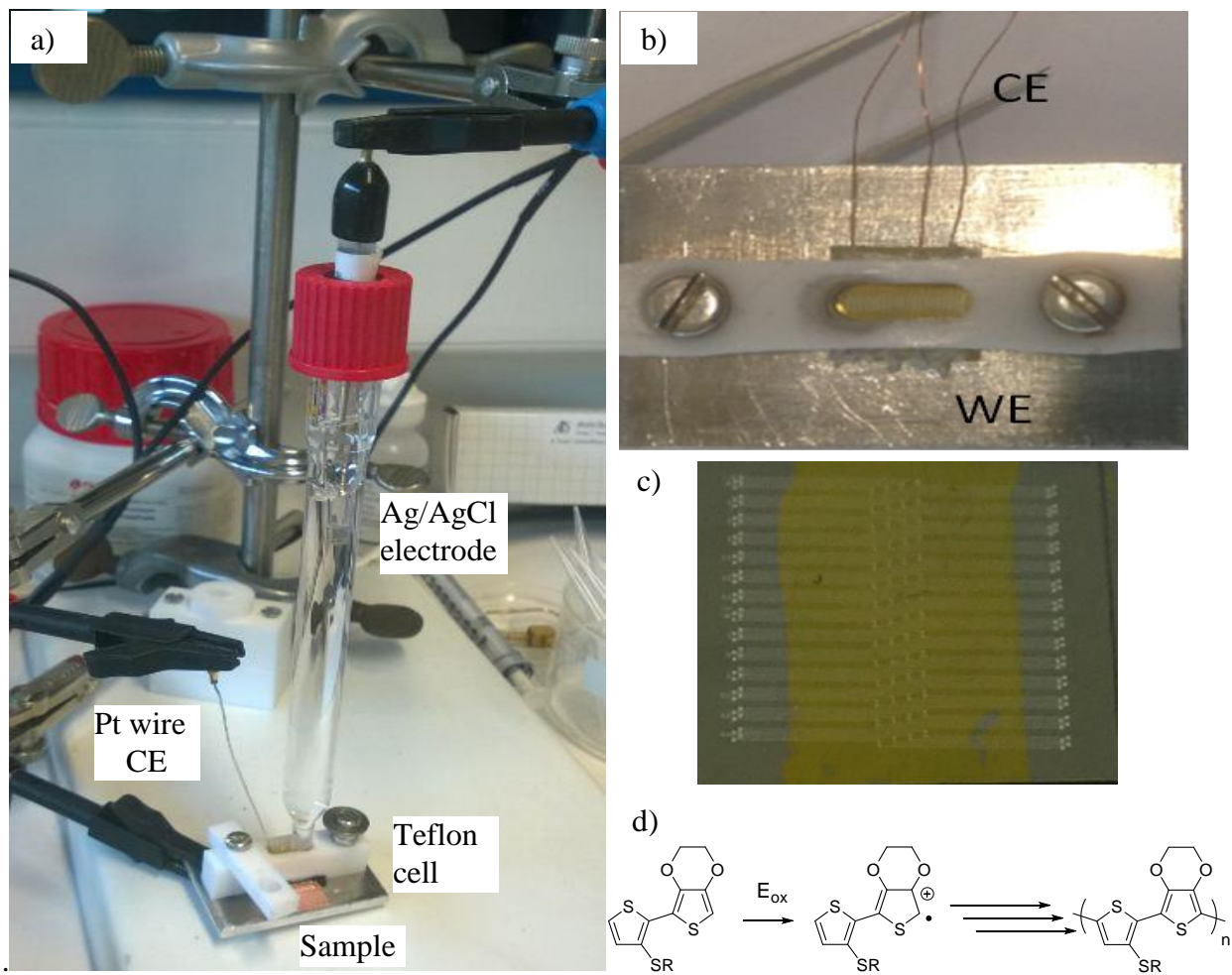

**Figure SI-6.** (a) Picture of the experimental setup for the electro-polymerization with the counter electrode (Pt wire) and Ag/AgCl reference electrode. (b) Picture of the lithographed electrodes covered by the Teflon cell containing the electrolyte solution. (c) Picture of the lithographed electrode after the electro-polymerization. (d) Reaction involved during the electro-polymerization.

## 5. NDR EFFECT IN BOTH POLARIZATIONS

The NDR (negative differential resistance) behavior was systematically and repeatedly observed for all the devices with different geometries, and for both positive and negative voltages (Figure SI-7). In this example, the maximum peak/valley ratio reaches a value around 17 in both polarities.

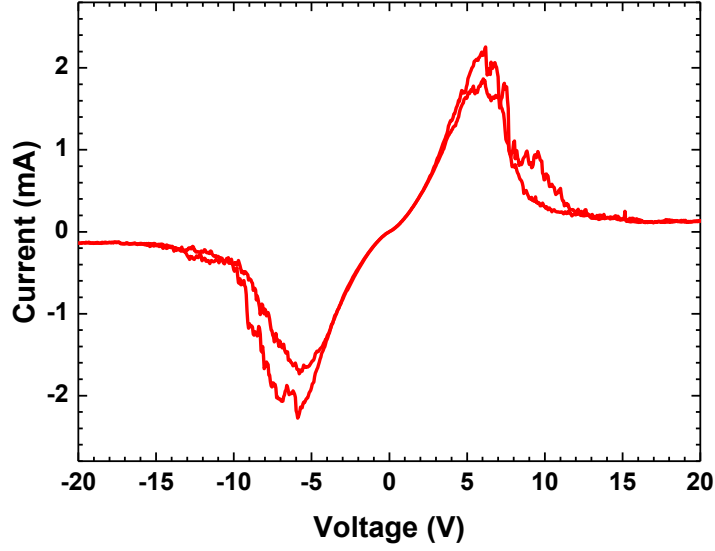

**Figure SI-7.** Current-voltage curves during a back and forth voltage sweeps, showing NDR effect in both polarizations, and measured on Formed-PTEDOT-AuNPs device with a length and width of 500 nm and 1 mm respectively for the electrode gap.

## 6. NDR AND MEMORY MECHANISMS

The hole capture by a NP is characterized by a capture time $\tau_C$ given by (according to Shockley-Hall-Read theory) [8]:

$$\tau_C = \frac{1}{\sigma_p v_p p} \qquad \text{Eq. SI-1}$$

with $\sigma_p$ the hole capture cross-section, $v_p$ the thermal velocity of charge carriers (holes here) in the polymers, p the hole density.

The hole detrapping time constant $\tau_E$ is related to the hole emission rate $e_p$ ($\tau_E=1/e_p$) with

$$e_p = \sigma_p v_p N_v \exp\left(-\frac{E_T}{kT}\right) \qquad \text{Eq. SI-2}$$

---

[8] W. Shockley and W.T. Read, Phys. Rev. 87, 835 (1952); Hall, R.N. (1951). Physical Review. 83 (1): 228

where $N_V$ is the density of states in the valence band (HOMO), k the Boltzmann constant, T the temperature and $E_T$ the energy level of NPs in the PTEDOT band gap (we consider the typical work function of Au NPs at 4.8-5 eV, and a HOMO of PTEDOT at ∼ 5.4 eV) - Fig. SI-8a. The results of the dynamic experiments (Fig. 7, main text) can be summarized as shown in Fig. SI-8b, in which we can distinguish the time domains and voltage windows of both capture and emission processes, with a region where both processes are in competition. In the zone where emission and capture occurs together, if we assume that the capture process dominates the emission, it may explain why emission is only measured for V between 5 and 8 V in the dynamic experiments described in Fig. 7b.

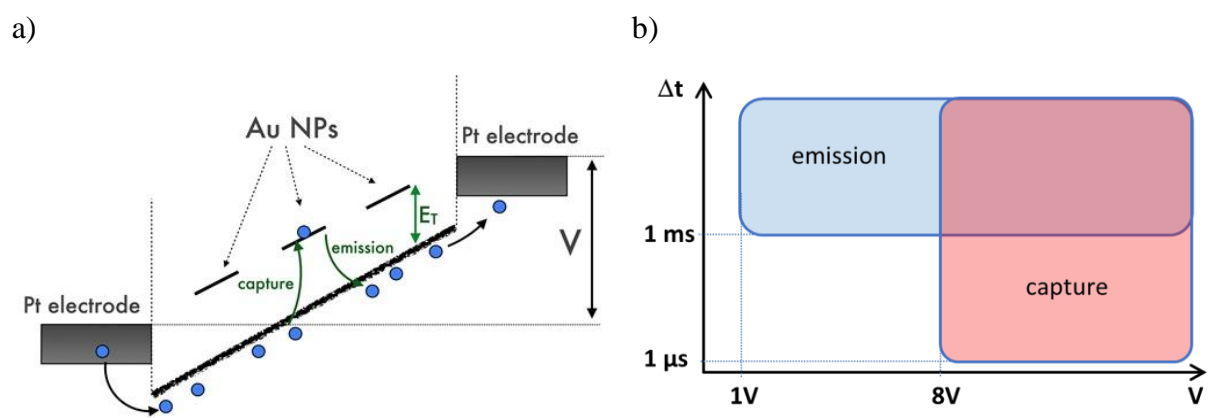

**Figure SI-8.** (a) Schematic energy diagram of the PTEDOT-AuNPs monolayer connected between two platinum electrodes. (b) Time constant - voltage window plot for the observation of emission and capture processes (numeric values from Fig; 7 in main text).